\documentclass[pra,twocolumn,a4paper,superscriptaddress,amsmath,amssymb,floatfix]{revtex4}
\usepackage{graphicx}
\usepackage{epstopdf}
\usepackage{braket}
\usepackage[colorlinks]{hyperref}
\hypersetup{%
   pdfstartpage=1,
   pdfpagemode=None,
   pdfstartview=FitH,
   colorlinks=true,
   linkcolor=blue
 }

\def\openone{\leavevmode\hbox{\small1\kern-3.3pt\normalsize1}}

\begin{document}

\author{J. Martin Berglund}
\affiliation{Theoretische Physik, 
  Universit\"{a}t  Kassel, Heinrich-Plett-Str. 40,
  D-34132 Kassel, Germany}

\author{Michael Drewsen}
\affiliation{Department of Physics and Astronomy,
  Aarhus University, Ny Munkegade 120, DK-8000 Aarhus, Denmark}

\author{Christiane P. Koch}
\email{E-mail: christiane.koch@uni-kassel.de}
\affiliation{Theoretische Physik, 
  Universit\"{a}t  Kassel, Heinrich-Plett-Str. 40,
  D-34132 Kassel, Germany}

\title{Femtosecond wavepacket interferometry \\ using the rotational
  dynamics of a trapped cold molecular ion}

\date{\today}

\begin{abstract}
  A Ramsey-type interferometer is suggested, 
  employing a cold trapped ion and 
  two time-delayed off-resonant femtosecond laser pulses.
  The laser light couples to the molecular polarization anisotropy, 
  inducing rotational wavepacket dynamics. 
  An interferogram is obtained from the delay dependent 
  populations of the final field-free rotational states. Current
  experimental capabilities for cooling and preparation of the initial
  state are found to yield an interferogram visibility of more than 80\%. 
  The interferograms can be used to determine the polarizability 
  anisotropy with an accuracy of about $\pm 2\%$, respectively $\pm
  5\%$, provided the uncertainty in the initial populations and
  measurement errors are confined to within the same limits. 
\end{abstract}

\maketitle

\section{Introduction}

Interference of waves both in the form of light and massive particles,
such as electrons, neutrals, and atoms, has proven to be a very
sensitive method of measuring physical quantities (See e.g.
Ref.~\cite{BermanBook} and
references therein). This includes precise measurements of
electromagnetic fields, gyromagnetic constants, gravitational
acceleration and rotation relative to an inertial system. When it
comes to interference of quantum objects with internal energy
structure, the method of Ramsey interferometry, involving two-level
objects, has been in particular successful~\cite{RamseyBook}. 
The accuracy in determining the internal state energies by Ramsey
interferometry scales inversely 
proportional to the square root of the interrogation time. Trapped
individual atomic ions, laser cooled or sympathetically cooled, 
have led to the most precise measurements to
date~\cite{BlattNature08}. Such  ions, 
occupying much less than a cubic micron, are ideal for controlled
laser excitations, and they can exhibit coherence time in the range of
hundreds of seconds~\cite{ChouPRL10}. 

\begin{figure}[tb]
  \centering
  \includegraphics[width=0.8\linewidth]{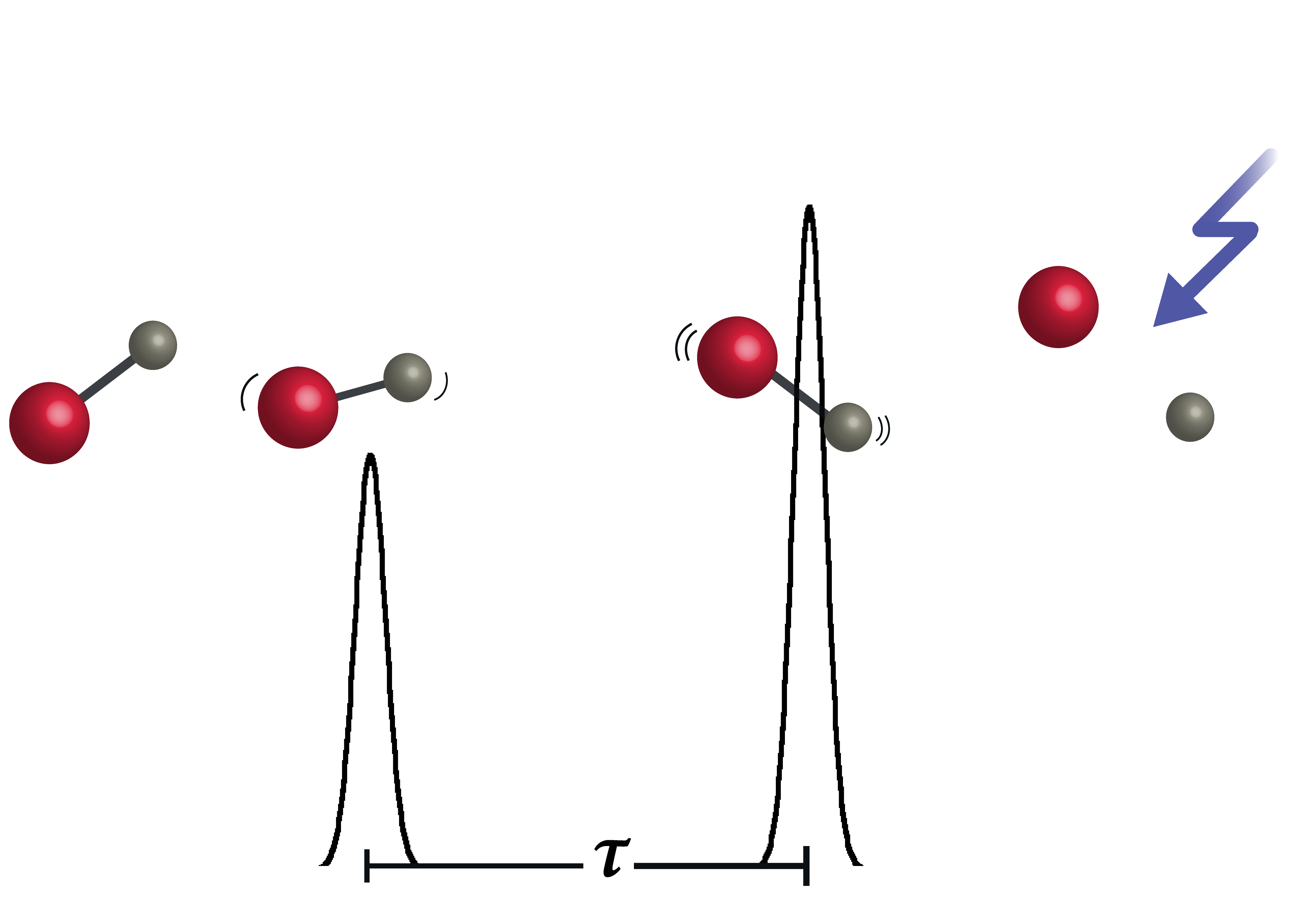}
  \caption{Schematic illustration of a rotational Ramsey
    interferometer with time evolving from left to right. A 
    sympathetically cooled MgH$^+$ molecular ion, 
    initially prepared in its ground  
    state, interacts with a first off-resonant laser
    pulse (the Mg$^+$ coolant ion is not shown). 
    Then the resulting wavepacket evolves freely for a controllable
    time delay $\tau$, until a second pulse is applied, generating a
    new wavepacket. This step is sensitive to the relative individual
    phases of the free-field states that make up the wavepacket prior
    to the second pulse and
    thus on the time delay. The final rotational populations are measured by a 
    state-sensitive dissociation pulse (blue arrow).}
  \label{fig:ramsey}
\end{figure}
Here a Ramsey-type interferometer using the 
rotational levels of a trapped and sympathetically cooled MgH$^+$
molecular ion~\cite{StaanumPRL08} is presented, as schematically 
illustrated in Fig.~\ref{fig:ramsey}. Compared to the standard version
employed in atom interferometry~\cite{HarocheBook}, 
the $\pi/2$ pulses are replaced by
off-resonant femtosecond laser pulses which couple to the
polarizability anisotropy of the molecule and induce rotational
wavepacket dynamics. Each field-free eigenstate in the wavepacket
acquires a specific phase during free evolution, equivalent to
specific optical path lengths acquired in light interferometry. The
resulting field-free populations are obtained from rotational state
selective photodisociation spectroscopy~\cite{Drewsen_diss} as  
a function of pulse delay. Their dependence on the time delay between
the pulses defines the interferogram.
The visibility of the interferogram depends on the laser pulse
parameters and the initial state. In particular, the visibility can be
enhanced by varying the intensity of the second pulse.

As an application of the  interferometer, the measurement of the
molecular polarizability anisotropy is discussed, taking both 
the experimental uncertainties of the initial populations
as well as population measurement errors into account. 
The interferometric method to determine the
static dipole polarizabilities discussed here
represents, for molecular ions, an interesting alternative to
microwave spectroscopy previously
used~\cite{TownesSchawlowBook,JacobsonPRA00}. 
Conversely, the interferometric technique may be used to probe local
electric fields, like the radio-frequency fields at the position of
the molecular ion in a (linear) Paul trap. The potentially very long
interogation times with trapped ions can make such an interferometer
extremely sensitive. 

The article is organized as follows: Section~\ref{sec:Model}
introduces the model and numerical tools. The results are presented in
Sec.~\ref{sec:Results}, starting with an assessment of the effective
rotor approximation in Sec.~\ref{subsec:ERA}, followed by an analysis
of the wavepacket created by the first laser pulse and the
characterization of the 
interferograms in Secs.~\ref{subsec:onepulse} and~\ref{subsec:IF}. The
prospects for measuring the polarizability anisotropy are discussed in
Sec.~\ref{subsec:alpha}. 
Section~\ref{sec:Conclusions} concludes. 

\section{Theoretical framework}
\label{sec:Model} 

A single trapped MgH$^+$ ion, translationally 
cooled down to sub-Kelvin temperatures and in its electronic ground
state, is considered.  
The ion interacts with femtosecond laser pulses which are far
off resonance from any transition in MgH$^+$ and  
linearly polarized along the laboratory fixed $z$-axis.
The Hamiltonian 
describing the rovibrational motion of the molecular ion and its 
interaction with the off-resonant field is given by ($\hbar = 1$)
\begin{equation}\label{eq:ham0}
  \mathbf{\hat{H}}_{2D} = \mathbf{\hat{T}}_r + V(\mathbf{\hat{r}}) 
  + \frac{\mathbf{\hat{J}}^2}{2m\mathbf{\hat{r}}^2}
  -\frac{I(t)}{2\epsilon_0c}\left(
    \Delta \alpha(\mathbf{\hat{r}}) \cos^2{\mathbf{\hat{\theta}}}
    + \alpha_{\perp}(\mathbf{\hat{r}})\right)\,,
\end{equation}
where the first two terms describe the radial kinetic and potential energy,
respectively. $\mathbf{\hat{J}}^2$ is the orbital 
angular momentum operator, and $I(t)$ the intensity profile of the
laser pulse. $\Delta\alpha(\mathbf{\hat{r}})$ denotes the molecular
polarizability anisotropy and $\alpha_{\perp}(\mathbf{\hat{r}})$ the
molecular polarizability perpendicular 
to the interatomic axis. $\theta$ is the angle between the
polarization vector of the laser pulse and the interatomic axis.
Assuming linear and parallel pulse polarizations, the 
Hamiltonian~(\ref{eq:ham0}) is independent of the azimutal angle
$\phi$. Therefore $\Delta m = 0$, and the $\cos^2{(\theta)}$-term
gives rise to the selection rule $\Delta j = 0,\pm 2$. 
The potential curve and polarizabilities are taken from
Ref.~\cite{AymarJPB09}. 

In Eq.~\eqref{eq:ham0}, the leading term of the light-matter
interaction is assumed to be via the ion's polarizability anisotropy.  
The interaction of the light with the permanent dipole moment of the
molecular ion has been neglected because it averages to zero if 
the laser pulses are off-resonant. 
Simulations of the vibrational dynamics under a pulse with
$\lambda_c \approx 800\,$nm, starting from the ground vibrational
level, yield a total
population of vibrationally excited levels of the order of $10^{-5}$
to $10^{-6}$, depending on the laser intensity. This confirms that 
practically no vibrational excitations take
place which is not surprising given that the 
energy difference between the ground and first excited vibrational
level corresponds to a wavelength of $\lambda \approx 6500\,$nm. It is
only for $v\approx 10$ that vibrational transitions may become
resonant. However, the matrix elements for these transitions is so
small that they play essentially no role. 

Since for low-lying rovibrational levels the vibrational
energy is much larger than the rotational energy, the two degrees of
freedom can be adiabatically separated~\cite{RosarioNJP09}. To this
end, the vibrational 
eigenfunctions are obtained by diagonalizing $\mathbf{\hat{H}}_{vib}$,
given by the first two terms in Eq.~\eqref{eq:ham0}. Denoting 
radial expectation values by $\braket{\cdot}_{\nu}$ for the 
${\nu}$th vibrational level, the vibrational motion can be integrated
out in Eq.~\eqref{eq:ham0}. This yields the so-called Effective Rotor
Approximation (ERA)~\cite{RosarioNJP09} where 
all $\mathbf{\hat{r}}$-dependent quantities in Eq.~(\ref{eq:ham0})
are replaced by their expectation values, 
\begin{equation}\label{eq:ham2}
  \mathbf{\hat{H}}_{\nu} = 
  B_{\nu}\mathbf{\hat{J}}^2
  -\frac{I(t)}{2\epsilon_0c}\left(\langle \Delta \alpha \rangle_{\nu}
    \cos^2{\mathbf{\hat{\theta}}} + \langle\alpha_{\perp}\rangle_{\nu}
  \right)\,,
\end{equation}
with $B_{\nu} = \frac{1}{2m}\braket{r^{-2}}_{\nu}$. 
The ERA neglects ro-vibrational couplings but 
goes beyond the rigid rotor approximation, since 
its parameters are obtained by integrating over the vibrational motion instead
of just replacing $\mathbf{\hat{r}}$ by the equilibrium 
distance. The values of the molecular parameters in
Eq.~\eqref{eq:ham2} for MgH$^+$, calculated using the \textit{ab
  initio} data of Ref.~\cite{AymarJPB09},  are listed in
Table~\ref{tab:rotpar}; they are found to be in good agreement with 
the experimental values of Ref.~\cite{BalfourCJP72}. 
\begin{table}[tb]
  \begin{centering}
    \begin{tabular}{|c| c| c|}
      \hline
      $B_{\nu=0}$ 
      & $\langle\Delta\alpha\rangle_{\nu=0}$
      & $\Braket{\alpha_{\perp}}_{\nu=0}$ \\\hline
      $6.3685\,$cm$^{-1}$  & $3.6634\cdot10^{-25}\,$cm$^{3}$ 
      & $7.3268\cdot10^{-25}\,$cm$^{3}$  \\
      \hline
    \end{tabular}
    \caption{Parameters of MgH$^+$, used in Hamiltonian~(\ref{eq:ham2}).}
    \label{tab:rotpar}
  \end{centering}
\end{table}

The rotational dynamics are characterized by the
rotational period, $\tau_{rot} = \hbar/(2B_{\nu=0})$, which amounts to 
$\tau_{rot} \approx 420\,$fs for MgH$^+$. This short rotational
period is a consequence of the large difference in the atomic masses and the 
small hydrogen mass. Rotational wavepacket revivals occur when
the wavepacket returns to its initial state. They can be analysed by the
correlation function, 
\begin{equation}\label{eq:corrfunct}
  C(t) = \Braket{\chi(0)|\chi(t)} = \sum_j |c_j|^2 e^{-iE_jt}\,,
\end{equation}
where $E_j= j(j + 1)B$ is the field-free rotational eigenenergy with
corresponding eigenfunction, 
\begin{equation}
  \chi^m_j(\theta) = \langle \theta|j,m\rangle= P^m_j(\theta)\,.
\end{equation}
Here, $P^m_j(\theta)$ is the associated Legendre function of degree $j$
and order $m$.  
Revivals occur at times $T$, for which $C(T) = C(0)$. That is, 
the conditions $2\pi k_j = E_jT$ with $k_j$ integer 
need to be fulfilled simultaneously for all $j$ which make up 
the wavepacket $\ket{\chi(t)}$.

The laser pulses $\varepsilon(t)$ which create the wavepackets
are assumed to have a 
Gaussian temporal envelope such that the pulse fluence becomes  
\begin{equation}\label{eq:int1}
  P(I_0,\tau_I) = 
  \frac{2}{\epsilon_0 c}\sqrt{\frac{\pi}{4\ln 2}}I_0 \tau_I\,.
\end{equation}
Here $I_0$ is the maximum pulse intensity and $\tau_I$ is the 
full width at half maximum (FWHM) duration of the intensity profile.
In particular, for constant fluence $P = P_0$, the intensity 
and pulse duration are inversely proportional.

Three different initial states will be considered in the
investigations presented below. Ideally for wavepacket interferometry,
the molecule is in its ground rotational state ($j=0$). In an
experiment, however, a completely pure initial state cannot be fully 
realized, but recent experiments with MgH$^+$ ions trapped in a
cryogenic environment have led to a nearly 80\% rotational ground state
population through helium buffer gas cooling~\cite{HansenNature14}.  
The ideal initial state is therefore compared to
a thermal ensemble with a rotational temperature of 20$\,$K and to an
incoherent ensemble prepared in current room-temperature
experiments by rotational
cooling~\cite{Drewsen} with the same ground state population 
($P_0\sim 0.38$) 
as a thermal ensemble at 20$\,$K. 
An incoherent initial state is described by a density operator, 
$\mathbf{\hat{\rho}}$:
\begin{equation}\label{eq:rho_0}
\mathbf{\hat{\rho}}(t=0) = \sum_{j=0}^{\infty}\sum_{m=-j}^j
a_j\Ket{j,m}\Bra{j,m}\,.
\end{equation}
For a thermal state at temperature $T$, 
$a_j = g_j\exp{(-\beta E_j)}/Z$  with 
$g_j=2j+1$, $\beta = 1/k_BT$, 
and $Z=\sum_j g_j \exp(-\beta E_j)$ the partition function, whereas 
for the experimentally prepared initial state, the values of $a_j$ are
taken from Ref.~\cite{Drewsen}. 
 
Since the timescale of the interferometer is much shorter than 
any decoherence time, the time evolution is coherent and the density
operator at time $t$ is given by
\begin{equation*}
\mathbf{\hat{\rho}}(t) = 
\mathbf{\hat{U}}(t)\mathbf{\hat{\rho}}(0)\mathbf{\hat{U}}^{\dagger}(t)\,.
\end{equation*}
Inserting Eq.~\eqref{eq:rho_0},
each $m$ state may be considered separately,
reducing the numerical effort in the calculations, 
since the Hamiltonian conserves $m$.
The time-dependent population of the state $j'$, with all corresponding  
$m$-states taken into account, is then obtained as 
\begin{equation}\label{eq:incoh}
  \rho_{j',j'}(t) = \sum_{j=0}^{\infty}a_j\sum_{m=j}^j
  \left|\Bra{j',m}\mathbf{\hat{U}}(t)\Ket{j,m}\right|^2\,,
\end{equation}
i.e., each pure state $\Ket{j,m}$ is propagated separately, using a
Chebychev propagator, and the
resulting population is added up incoherently with its proper
weight. The summation over the initially populated values of $j$ 
in Eq.~(\ref{eq:incoh}) can be truncated at $j^{ini}_{max} = 6$. 
The basis set expansion in the Legendre polynomials is found to be
converged for $j_{max} = 20$, provided $I_0 \le 4 \times 10^{13}$ W/cm$^2$.

\section{Results}
\label{sec:Results}

Before presenting the results obtained with
Hamiltonian~\eqref{eq:ham2}, the accuracy of the ERA is checked in
Sec.~\ref{subsec:ERA}. 
Then the interferometer is analyzed in a step-wise fashion, starting
in Sec.~\ref{subsec:onepulse} with the dependence of the wavepacket,
that is created by the first pulse, on pulse intensity and duration. 
The complete interferometer time evolution is presented in
Sec.~\ref{subsec:IF}, determining pulse parameters that yield
high-visibility interferograms. Prospects for using the interferometer
to measure the molecular polarizability are discussed in
Sec.~\ref{subsec:alpha}. 

\subsection{Accuracy of the effective rotor approximation}
\label{subsec:ERA}

The ERA, Eq.~\eqref{eq:ham2}, is tested against the full rovibrational
dynamics, generated by Hamiltonian~\eqref{eq:ham0}, considering the 
interaction with one pulse of $100\,$fs duration. To this end, 
the radial coordinate in Eq.~\eqref{eq:ham0} is represented on a
Fourier grid, and the time-dependent Schr\"odinger equation with
Hamiltonian~\eqref{eq:ham0} is solved. 
The absolute difference in the final-time population is found to be
within 0.01 for $j=0,\ldots,6$ and intensities $I_0\le 1\times
10^{13}\,$W/cm$^2$. The relative error amounts to less than one
percent for $j$ up to $j=4$ and intensities  
up to $1\times10^{13}$ W/cm$^2$. For intensities 
$1\times 10^{13}\,$W/cm$^2\le I_0 \le 4\times 10^{13}\,$W/cm$^2$, the
absolute error due to the ERA is within 0.015 for $j=0,\ldots,6$,
whereas the relative error reaches 10\% to 15\%.
While both absolute and relative errors become larger for higher $j$
states, the ERA is applicable for our purposes since  
low-lying $j$ states ($j \le 6$) are most relevant for interferometry
and only  moderate pulse intensities will be considered to ease
experimental feasibility. 

\subsection{Creating a rotational wavepacket by a single
  femtosecond laser pulse} 
\label{subsec:onepulse}

\begin{figure*}[tb]
  \centering
  \includegraphics[width=\linewidth]{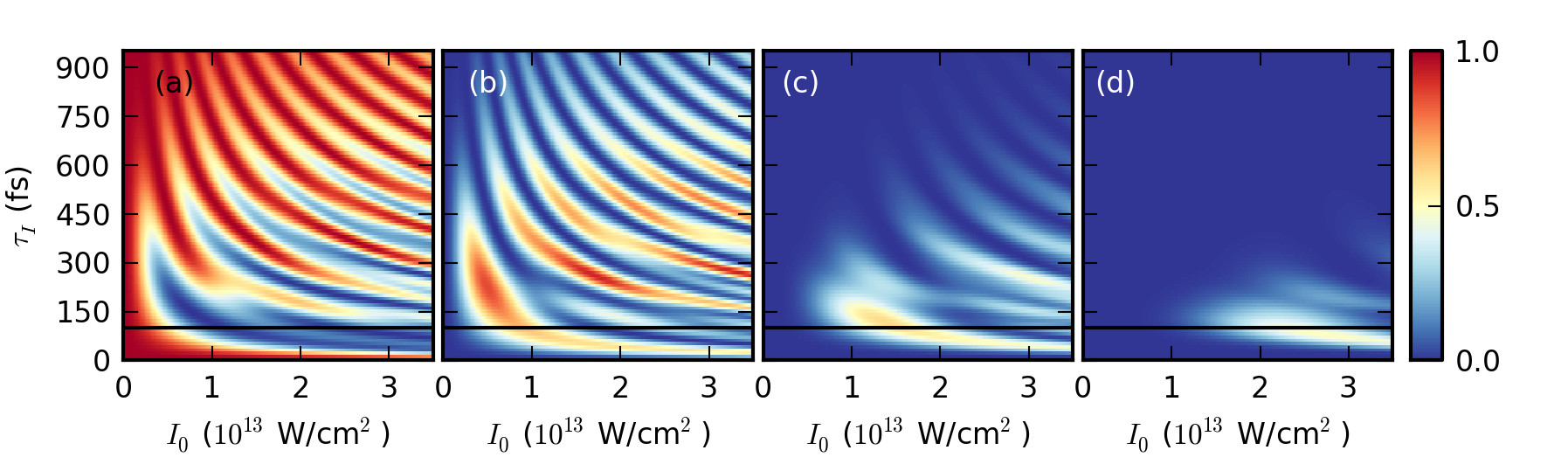}
  \caption{Final population of field-free rotational states 
    (a:  $j = 0$, b: $j = 2$, c: $j = 4$, d: $j = 6$)
    after interaction with a single laser 
    pulse as function of pulse intensity and duration. The initial 
    state is $j=0$. The black line indicates pulses of 100$\,$fs FWHM.}
  \label{fig:landscape1}
\end{figure*}
\begin{figure}[tb]
  \centering
  \includegraphics[width=0.95\linewidth]{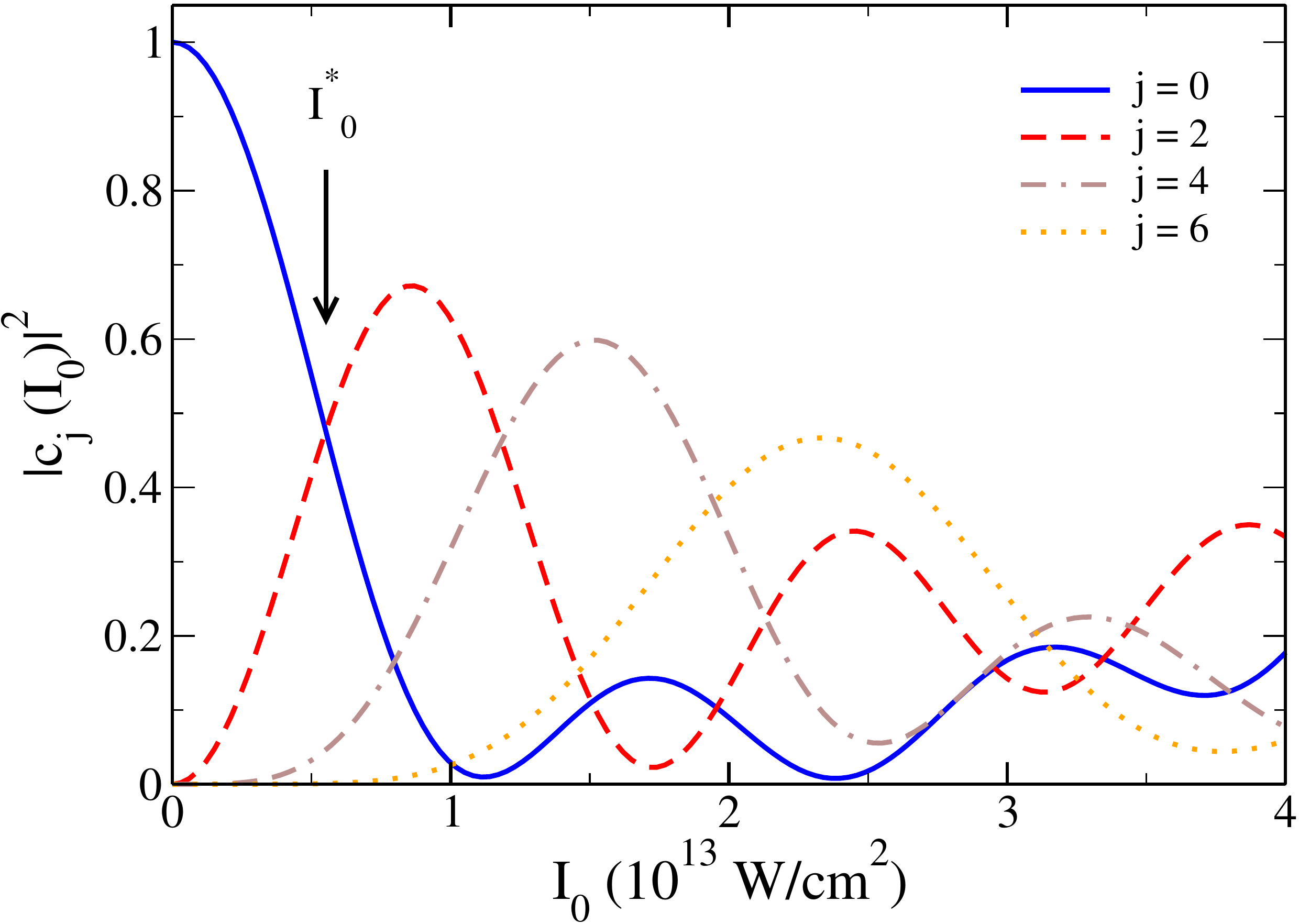}
  \caption{Final population of field-free rotational states 
    after interaction with a single 100$\,$fs pulse as a function of
    pulse intensity. The arrow at $I^*_0$  
    indicates the laser intensity for which equal population of the states 
    $j=0$ and $j=2$ is obtained, corresponding to the black line in
    Fig.~\ref{fig:landscape1}.} 
\label{fig:op}
\end{figure}
Starting from a pure initial state ($j=0$, $m=0$ ), a femtosecond
laser pulse creates a rotational wavepacket which, due to the
selection rules, is made up of states $j=0,2,4,\ldots$, all with $m=0$. 
Since the laser-molecule interaction is off-resonant and Gaussian
pulse envelopes are assumed, the composition
of the wavepacket is only determined by the intensity and duration of
the pulse. The dependence of the final rotational state populations
on pulse intensity and duration is shown in Fig.~\ref{fig:landscape1} 
for $j=0,2,4,6$.
Curves of constant pulse duration (intensity) correspond to horizontal
(vertical) cuts in Fig.~\ref{fig:landscape1}.  
The black line  indicates pulses of $\tau_I=100\,$fs
duration. Constant pulse fluences correspond to hyperbolas in the 
landscape, cf. Eq.~\eqref{eq:int1}. Hyperbolas are clearly visible in
Fig.~\ref{fig:landscape1}, indicating that it is the pulse fluence
that determines the population transfer. 
For the pulse parameters corresponding to the upper left part of each
panel in Fig.~\ref{fig:landscape1}, 
the molecule can approximately be described as a two-level
system, consisting of the states $j=0$ and $j=2$. This should allow
for the closest analogy to a Ramsey interferometer as used with
atoms~\cite{HarocheBook}. 
For short pulses, $\tau_I \lesssim 300$ fs, the final populations show a
more complicated behavior, with more states being significantly
populated, save for very small $I_0$.
A wide range of pulse parameters gives rise to significant population
of $j=2$, see the lower left corner of Fig.~\ref{fig:landscape1} (b).
Significant population of the $j=4$ state is obtained for intensities
larger than  $0.8 \times 10^{13}\,$W/cm$^2$ and 
pulses shorter than 150$\,$fs. However, the $j=4$ state starts to be
populated already at smaller pulse intensities. This inhibits a perfect
50\%-50\% superposition  of the states $j=0$ and $j=2$.

In a typical experimental setup, the transform-limited pulse
duration is fixed, i.e., to 100$\,$fs, whereas the intensity is more
easily varied. The dependence of the final state populations on pulse
intensity for a 100$\,$fs pulse is presented in Fig.~\ref{fig:op}, 
as marked by the black lines in Fig.~\ref{fig:landscape1}. For low
intensity, the behaviour is similar to a two-level system, as to be
expected from the nearest neighbour coupling in
Hamiltonian~\eqref{eq:ham2}. As the intensity increases, more levels
are populated and the result deviates more and more from the simple
two-level picture. For even higher pulse intensities,   
recurring peaks of low $j$ states appear. Note that only a small
number of levels can be superimposed at a given field 
intensity. Intensities for which population curves cross,
indicating equal population, are particularly interesting for
interferometry since they should yield good contrast. The first such
occurrence, at $I_0 \approx 0.5 \cdot 10^{13}$ W/cm$^2$ for $j =0$ and
$j=2$, is marked by $I^*_0$ in the figure. Another crossing occurs at 
$I_0 \approx 1.1 \cdot 10^{13}$ W/cm$^2$ for $j=2$ and $j=4$. 

\begin{figure}[tb]
  \centering
  \includegraphics[width=0.95\linewidth]{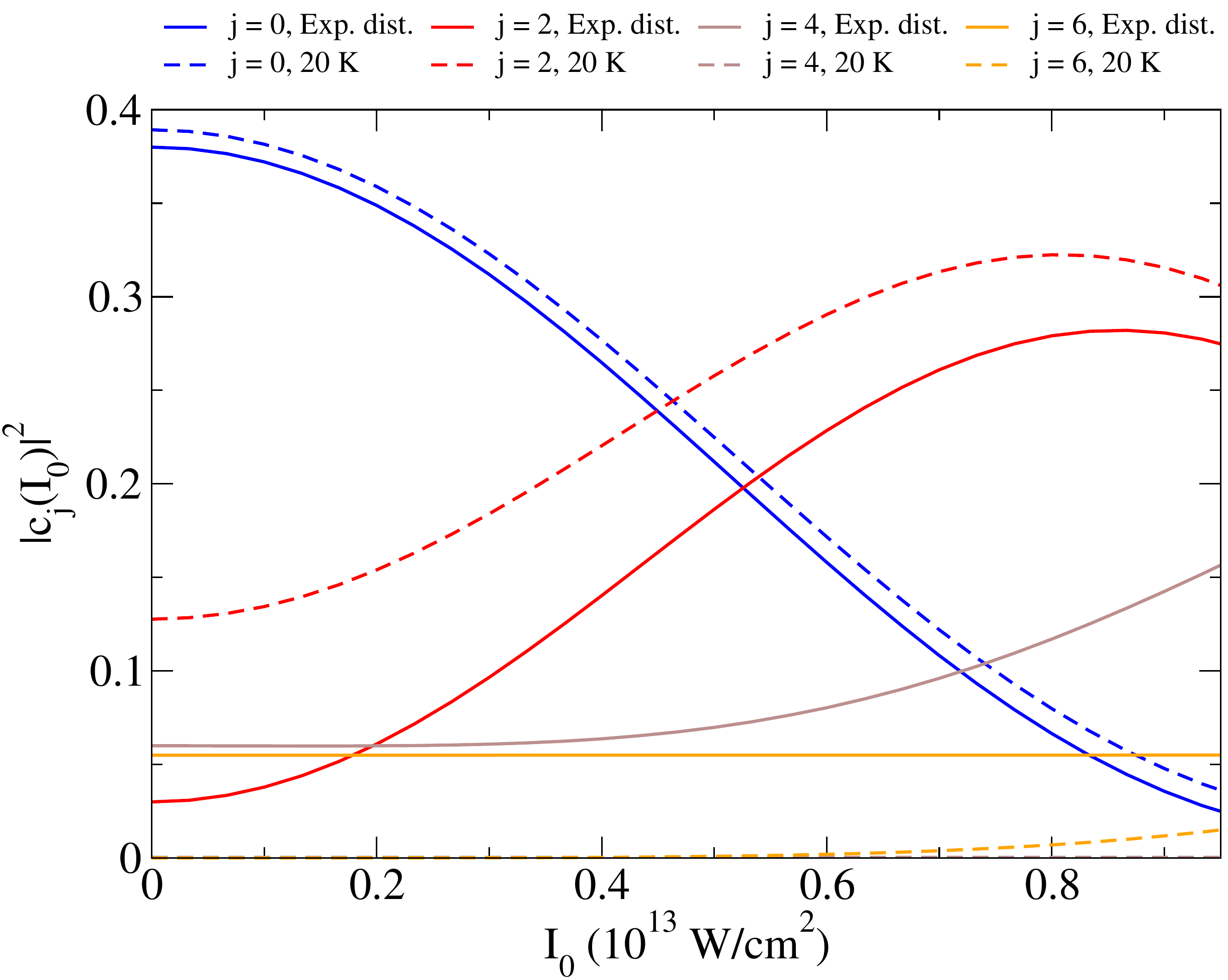}
  \caption{Same as Fig.~\ref{fig:op} but for incoherent initial
    states, corresponding to the experimental distribution of
    Ref.~\cite{Drewsen} (solid lines) and a thermal ensemble at 20$\,$K
    (dashed lines).} 
  \label{fig:ensemble}
\end{figure}
The dynamics becomes more involved for incoherent initial states 
since in this case also states with $m \neq 0$ are initially
populated. Figure~\ref{fig:ensemble} shows the final field-free state
populations  as a function of laser intensity for a 100$\,$fs pulse,
comparing two different initial states, 
the experimental distribution of Ref.~\cite{Drewsen} and a 
thermal ensemble at 20$\,$K. The final $j=0$ population of the two
ensembles behaves very similarly, with only a small offset in 
initial population. Also, both ensembles qualitatively lead to the
same dynamics as the pure initial state in Fig.~\ref{fig:op},
confirming that the ensemble dynamics is dominated by $j=0$, at least
for the intensities examined in Fig.~\ref{fig:ensemble}. 
The final $j=2$ population on the other hand shows some 
differences between the two initial ensembles, with the peak occurring
slightly earlier and the maximum population difference being slightly
smaller for the thermal ensemble. For $j=4$ and $j=6$, the initial
populations are negligible in the thermal ensemble and small,
but non-zero in the experimental distribution of
Ref.~\cite{Drewsen}. Therefore the resulting final 
populations after interaction with the pulse in
Fig.~\ref{fig:ensemble} are similar to those obtained for the pure
initial state in Fig.~\ref{fig:op}. Although the initial population in
$j=4$ and $j=6$ is non-negliglible for the experimental distribution, these
states do not take part in the dynamics for pulse intensities up to
$0.5\times 10^{13}$ W/cm$^2$. This is promising in view of obtaining
high-visibility interferograms even with incoherent initial states. 

\subsection{Creating and probing rotational wavepackets using a
  sequence   of two  femtosecond laser pulses}
\label{subsec:IF}

\begin{figure}[tb]
  \centering
  \includegraphics[width=0.9\linewidth]{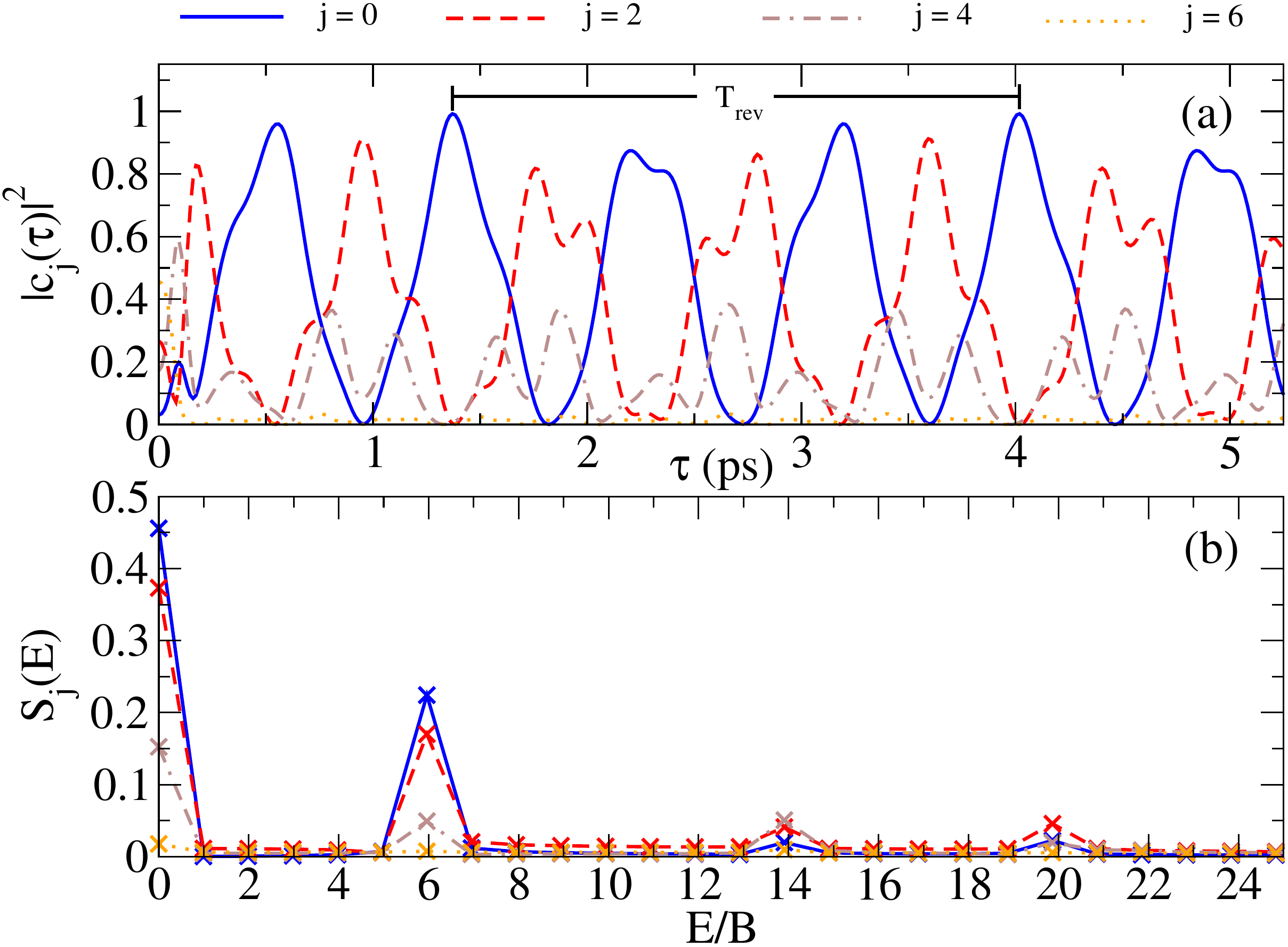}
  \caption{a: Interferogram, i.e., the final  populations of the 
    field-free rotational states, obtained after interaction of the
    molecule with two laser pulses, as a function of time delay. 
    Both pulses have $I_0 \approx 0.5 \times 10^{13}$ W/cm$^2$ 
    (as indicated by the arrow in Fig.~\ref{fig:op}) 
    and $\tau_I = 100\,$fs.
    b: Spectrum $S_j$ of the populations shown in panel (a).}
  \label{fig:interfer}
\end{figure}
Interferograms are obtained when the molecule interacts with two
laser pulses, separated by a time delay, $\tau$. 
For simplicity, the parameters $I_0$ and $\tau_I$ are chosen to be the
same for both pulses except where indicated.
First, consider the pure initial state with $j = 0$.
The laser intensity and pulse duration were chosen such that the first
pulse yields equal populations for $j=0$ and $j=2$,  as 
marked by $I^*_0$ in Fig.~\ref{fig:op}. 
Since very little population is transferred to states other than
$j=0$ and $j=2$ by the first pulse
($|c_0|^2 = |c_2|^2 \approx 0.47$, $|c_4|^2 \approx 0.06$), 
a simple interference pattern is obtained in
Fig.~\ref{fig:interfer} with the population predominantly in $j=0$ and
$j=2$ for all time delays. 
The condition fo revivals of the wavepacket created by the first pulse 
becomes $T_{rev} = \pi / B \approx 2.6\,$ps. 
The revival time is indicated in Fig.~\ref{fig:interfer}, confirming
the estimate of 2.6$\,$ps predicted by the correlation function.
A very high contrast, or, equivalently, large visibility, is observed 
for the lower  $j$ states. The visibility is defined as
\begin{equation}\label{eq:vis}
  V_j = \frac{|c_{j,max}|^2 - |c_{j,min}|^2}{|c_{j,max}|^2 + |c_{j,min}|^2}\,,
\end{equation}
where $|c_{j,max}|^2$ ($|c_{j,min}|^2$) is the maximum (minimum)
population of the $j$th state.
The $j=0$ state reaches an almost perfect visibility of one and the 
$j=2$ state around 0.9, reflecting the almost perfect 50\%-50\%
superposition of these two states. 

The Fourier spectra of the delay-dependent final populations in 
Fig.~\ref{fig:interfer}(a) are shown in Fig.~\ref{fig:interfer}(b), 
$S_j = \sqrt{\mathcal{F}[f_j]}$ with $\mathcal{F}$ denoting the
Fourier transform and $f_j(\tau) = |c_j(\tau)|^2$. As is evident from
Fig.~\ref{fig:interfer}(b), the spectrum is useful to visualize 
the components of the wave packet created by the first pulse, since it
displays peaks at the eigenenergies ($E_0=0$, $E_2=6B$, $E_4=20B$)
as well as at the quantum beats ($E_{2/4}=14B$). The similar peak
heights of $S_0$ and $S_2$ at $E_0=0$, $E_2=6B$ reflect the almost
identical population of the states $j=0$ and $j=2$.  
They differ at higher energies, since population from the  $j=2$ state
is further excited into the $j=4$ state.

\begin{figure}[tb]
  \centering
  \includegraphics[width=0.9\linewidth]{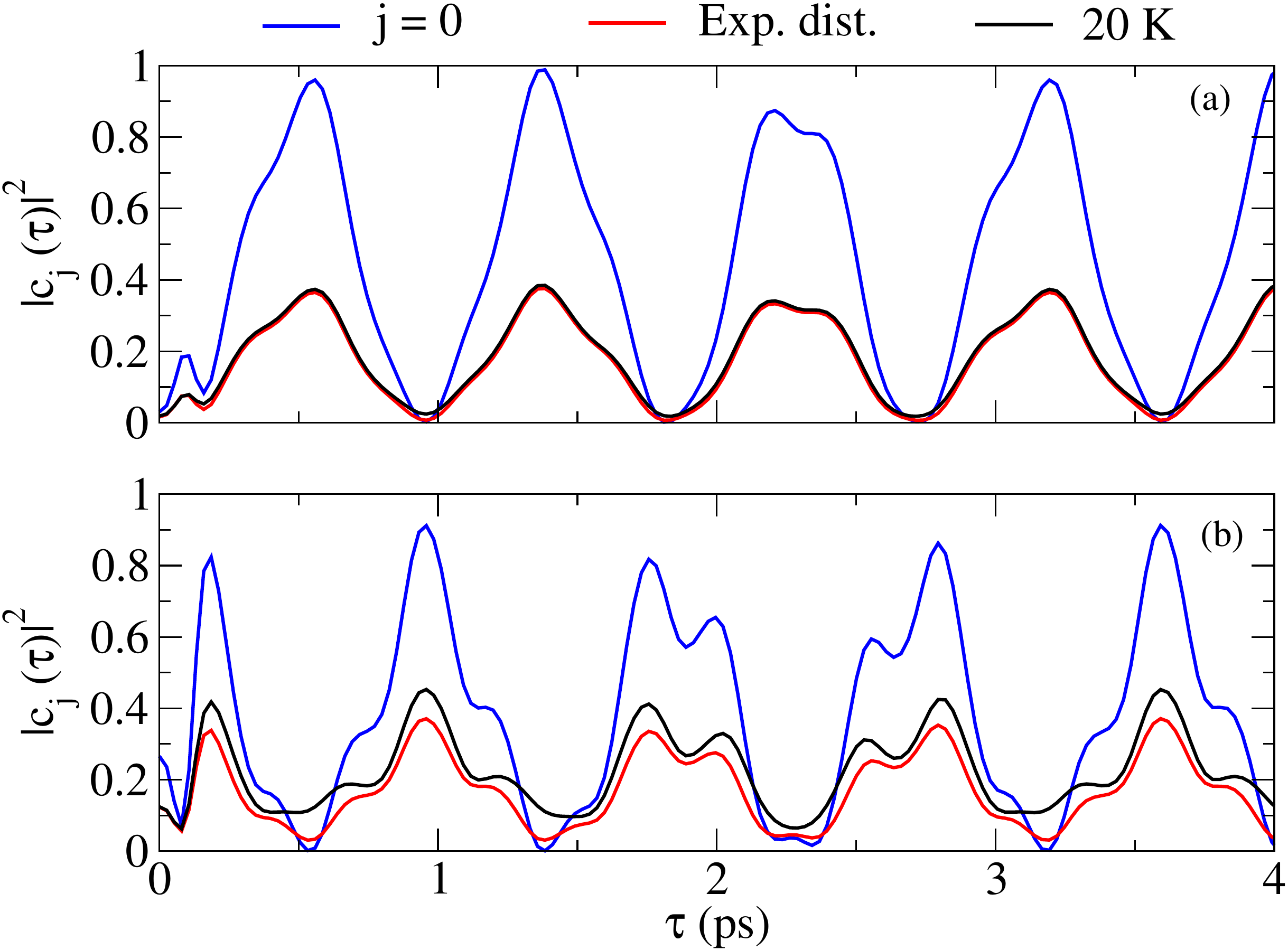}
  \caption{Interferograms for incoherent initial states, corresponding
    to the experimental distribution of Ref.~\cite{Drewsen} (red) and
    a thermal ensemble at 20$\,$K (black), compared to that of a pure
    initial state with $j=0$ (blue):  
    Final populations of the field-free rotational states (a: $j=0$,
    b: $j=2$) as a function of pulse delay. Pulse parameters as in
    Fig.~\ref{fig:interfer}.} 
  \label{fig:incoh}
\end{figure}
Next, Fig.~\ref{fig:incoh} examines the potentially detrimental effect
of incoherence in the initial states on the interferogram. It compares
the interference patterns for the pure initial state of
Fig.~\ref{fig:interfer} with those obtained for the 
experimental distribution of Ref.~\cite{Drewsen} and a 20$\,$K
thermal ensemble. When measuring the final population of $j=0$, 
the two incoherent ensembles give practically identical results, cf. 
Fig.~\ref{fig:incoh}(a). Their delay-dependence is qualitatively the
same as that of the pure initial state, but the maximum amplitudes are  
reduced from close to one down to about $0.4$. Notwithstanding, 
the visibility, $V_0$, is only reduced to 0.96 for the experimental
distribution and 0.91 for the thermal ensemble. 
For $j=2$ (Fig.~\ref{fig:incoh}(b)), the 
maximum population still amounts to about $0.4$, at least for the
experimental distribution. The interferograms of the incoherent ensembles 
differ significantly from each other, and they also differ
qualitatively from the interferogram of the pure initial state. This
simply reflects the fact that the dynamics of the $j=2$ state is
affected by more states in the initial ensemble. If one wants to test
the composition of the initial state interferometrically, measurement
of $j=2$ is therefore preferred to $j=0$. 
The visibility of the interferograms for $j=2$ is reduced, compared to
the pure state, to 0.85 for the experimental distribution of
Ref.~\cite{Drewsen} and to 0.76 for the thermal state. These numbers
are very encouraging in view of the feasibility of a rotational
interferometer. In summary, incoherence in the initial state does not
preclude interferometry, in particular if one measures the $j=0$
population. 

\subsection{Prospect of measuring the polarizability
  anisotropy} 
\label{subsec:alpha}

\begin{figure}[tb]
  \centering
  \includegraphics[width=0.99\linewidth]{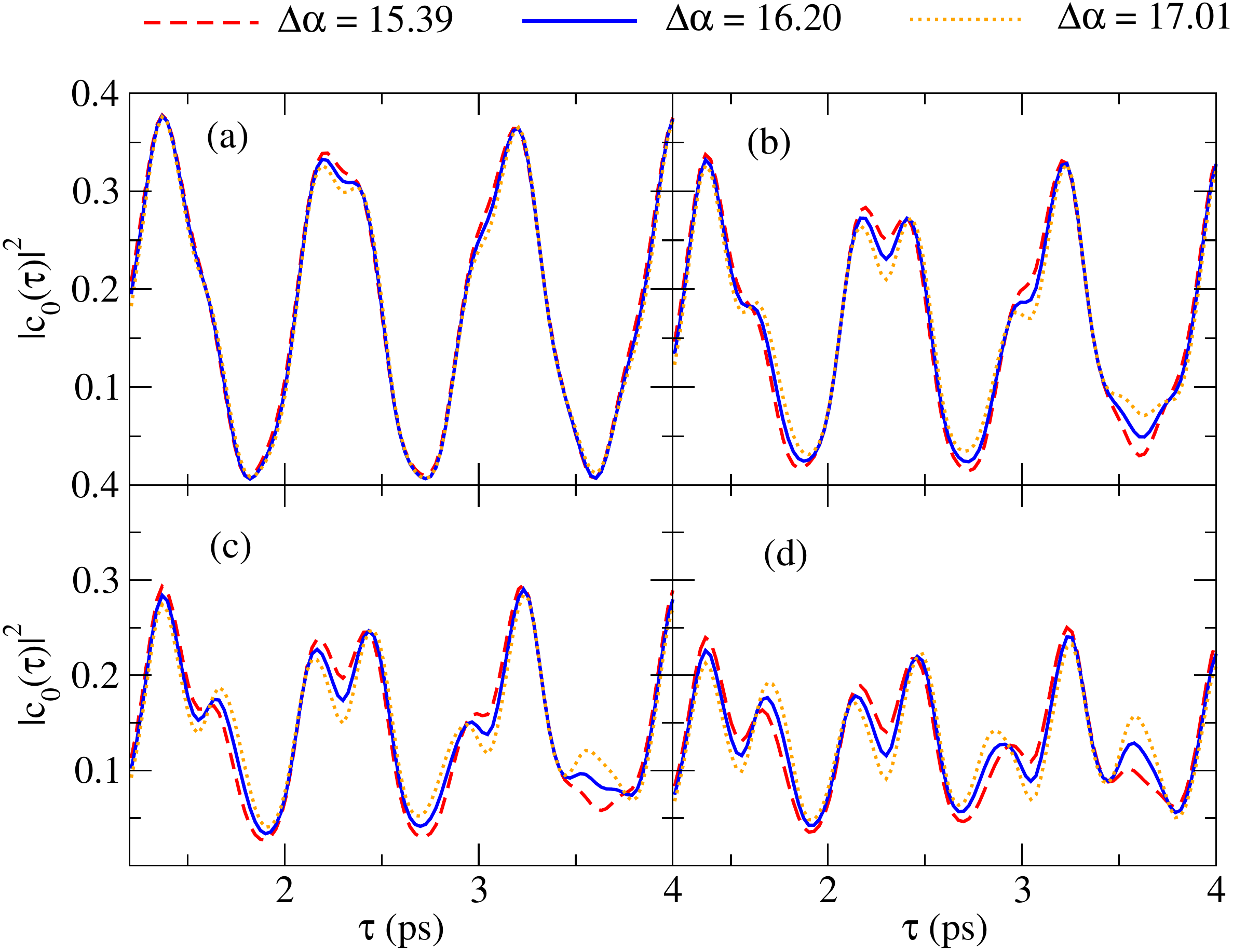}
  \caption{Interferograms, i.e., final $j=0$ populations as function
    of delay, for different values of the polarizability
    anisotropy $\Delta\alpha$ and increasing intensity of the second
    pulse (a: identical pulse intensities, b: $I^2_{0}=1.4I^1_{0}$, 
    c: $I^2_{0}=1.6 I^1_{0}$, d: $I^2_{0}=1.8I^1_{0}$). 
    The pulse duration is $\tau_I = 100\,$fs for both pulses, and 
    the intensity of the first pulse is 
    $I_0^1 = 0.55 \times 10^{13}\,$W/cm$^2$. The initial state
    corresponds to the experimental distribution of
    Ref.~\cite{Drewsen}.}
\label{fig:dalpha_pr}
\end{figure}
The high visibility of the interferograms presented in
Sec.~\ref{subsec:IF} suggests that the rotational interferometer can be
employed to determine molecular parameters such as the polarizability
anisotropy. Figure~\ref{fig:dalpha_pr} shows interferograms obtained by
measuring the final  $j=0$ population for several values of
$\Delta\alpha$ -- in atomic units: 16.20$\,$a$_0^3$ (the \textit{ab initio}
value), 17.01$\,$a$_0^3$ (5\% larger) and 15.39$\,$a$_0^3$ (5\%
smaller), increasing the peak intensity of the second pulse. Whereas the
interferograms are essential identical for the three values of the
polarizability anitropy if the pulses have the same intensity
(Fig.~\ref{fig:dalpha_pr}a), the curves become more and more
distinguishable when the intensity of the second pulse is
increased (Fig.~\ref{fig:dalpha_pr}b-d). The better distinguishability
comes at the price of a slightly deteriorated visibility of 0.79 for 
$I^2_0=1.6 I_0^1$ (Fig.~\ref{fig:dalpha_pr}c)
compared to 0.97 for equal intensities  (Fig.~\ref{fig:dalpha_pr}a). 
Particularly promising features are observed for $I^2_0=1.6 I_0^1$ in 
Fig.~\ref{fig:dalpha_pr}c for delays around $1.2\,$ps, $2.3\,$ps,
$3\,$ps and 
$3.5\,$ps. Such clear differences should still be observable, even
when inevitable error bars are taken into account. 

\begin{figure}[tb]
  \centering
  \includegraphics[width=0.99\linewidth]{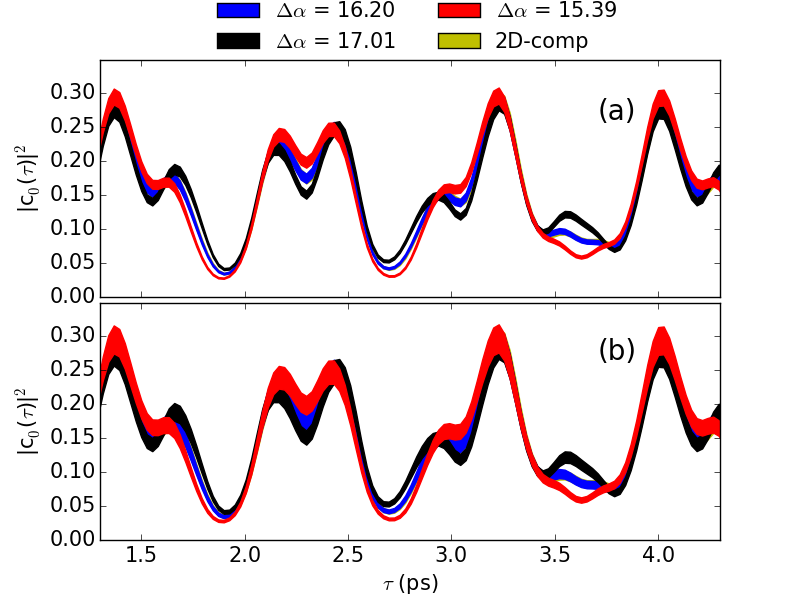}
  \caption{Interferograms, i.e., final $j=0$ populations as a function
    of time delay, for different values of the polarizability
    anisotropy $\Delta\alpha$, accounting for
    2\% (a) and 5\% (b) measurement uncertainty in the final
    populations. 
    The initial state corresponds to the experimental distribution 
    of Ref.~\cite{Drewsen} with 2\% uncertainty in the initial
    populations taken into account ($\tau_I= 100\,$fs for both pulses,
    $I_0 = 0.55\times 10^{13}\,$W/cm$^2$ for the first pulse, 
    $I_0 = 1.6 \cdot 0.55\times 10^{13}\,$W/cm$^2$ for the second pulse).}
\label{fig:dalpha1}
\end{figure}
\begin{figure}[tb]
  \centering
  \includegraphics[width=0.99\linewidth]{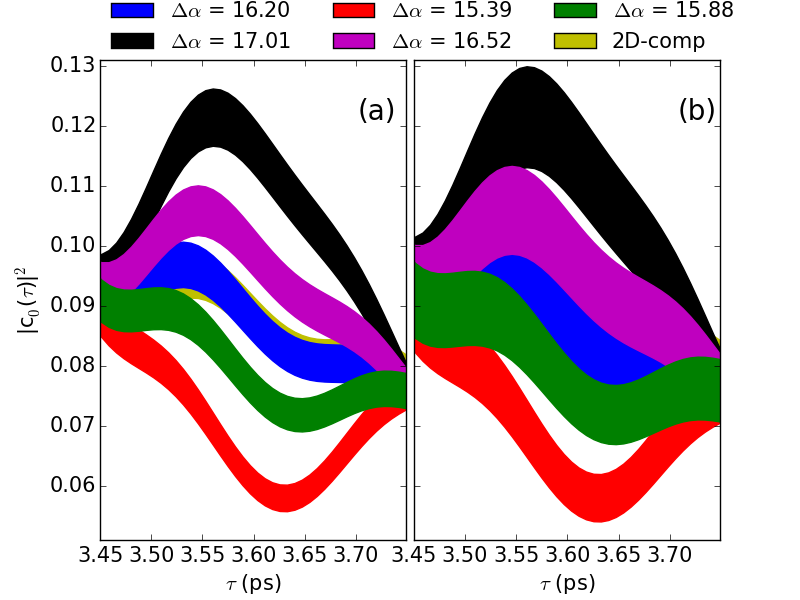}
  \caption{Same as Fig.~\ref{fig:dalpha1} for a smaller range of pulse
    delays, $\tau \in [3.45\,$ps$,3.75\,$ps$]$.}
\label{fig:dalpha_zoom}
\end{figure}
Measurement errors in the final $j=0$ population of 2\% and 5\% are
assumed in Fig.~\ref{fig:dalpha1}(a) and (b), respectively.
In the case of a 2\% measurement error, the interferograms  are easily
distinguishable from each other, i.e., the curves including error bars
do not overlap, in various ranges of time delays, 
for example for $\tau\in[1.5\,$ps,2.0$\,$ps$]$, $[2.7\,$ps,2.9$\,$ps$]$, or 
$[3.5\,$ps,3.8$\,$ps$]$. For a 5\% measurement error, 
however, only the latter interval of pulse delays allows to determine 
the polarizability anisotropy with a confidence level of $\pm$5\%. 
Within this region the interferogram is particularly sensitive to small
variations in the molecular polarizability anisotropy. A zoom of this
region is shown in Fig.~\ref{fig:dalpha_zoom}, where two more values of 
$\Delta\alpha$, 16.52$\,$a$_0^3$ and 15.88$\,$a$_0^3$, larger by $\pm$2\% than the
\textit{ab initio} value, have also been included. 
It is seen that for a 2\% measurement error the interferometer is 
readily sensitive to $\pm 2\%$ shifts in $\Delta\alpha$, whereas for a
5\% measurement error, only shifts of $\pm 5\%$ can unequivocally be
distinguished. 

Our predictions for the sensitivity of the rotational interferometer 
are based on averaging over many wavepacket calculations to account
for inevitable experimental inaccuracies. One might argue that the
corresponding noise effects may come into play differently in the
ERA and the full rovibrational dynamics. In order to be sure that our
conclusions are not compromised by a break-down of the ERA, 
Figs.~\ref{fig:dalpha1} and~\ref{fig:dalpha_zoom}
compare the interferogram obtained within the ERA for the \textit{ab
  initio} value of $\Delta\alpha$ with that obtained from full
two-dimensional calculations, using Hamiltonian~\eqref{eq:ham0}. While
slight deviations in the error bars between ERA and 2D model are
visible, in particular in Fig.~\ref{fig:dalpha1}(b)
and~\ref{fig:dalpha_zoom}(b), they are sufficiently small not to
affect the confidence levels stated above. That is, a sensitivity of the
interferometer to changes in the polarizability anisotropy of $\pm$2\%
($\pm$5\%) requires the measurement errors not to exceed the same
level.

\section{Conclusions}
\label{sec:Conclusions}

A Ramsey-type interferometer, employing off-resonant femtosecond laser
pulses to induce rotational wavepacket dynamics
in a trapped, cooled MgH$^+$ molecular ion, can be implemented using
current experimental capabilities. Unlike in atom interferometry, where
it is comparatively straight-forward to pick two isolated levels for 
Rabi cycling, application of the second pulse leads to rotational
ladder climbing in the molecule. Perfect visibility of the 
interferogram can thus only be obtained for $j=0$, whereas 
measuring $j=2$ leads
to 90\%. It also requires a pure initial state with $j=0$, $m=0$.  

Preparing a molecule perfectly in its rovibrational ground state is a
very challenging task. However, a ground state population of 38\%, as
prepared in Ref.~\cite{Drewsen}, is
found to decrease the visibility for $j=0$ to only 96\% and that for
$j=2$ to only 85\%. Even for a thermal distribution with a rotational
temperature of 20$\,$K, high-visibility interferograms are predicted. 
The required intensities are moderate for the case of the MgH$^+$
molecular ion, of the order of $10^{12}\,$W/cm$^2$ for 100$\,$fs laser
pulses. This suggests feasibility of a rotational Ramsey
interferometer, combining standard trapping and cooling techniques for
molecular ions with 800$\,$nm femtosecond laser pulses.

Such an interferometer could be used for example to determine the
molecular polarizability anisotropy in the vibrational ground state 
by comparing an experimental interferogram to theoretical predictions
for various values of $\langle\Delta\alpha\rangle_0$. Taking  
experimental uncertainties in the initial populations as well as
population measurement errors into account, the interferometer is
found to be sensitive to changes in the polarizability anisotropy of 
$\pm 2$\%, respectively, $\pm 5$\%
assuming the same level of experimental inaccuracy.

It will be interesting to see whether the rotational Ramsey
interferometer can also be used to determine the dependence of the
molecular polarizability anisotropy on the internuclear separation. A
possible route could be provided by recording rotational
interferograms for several vibrational states. Alternatively, one
could exploit the full rovibrational dynamics. In both cases, however,
it might turn out to be difficult to disentangle effects that are due
to the shape of the potential energy curve (which is also known only
approximately) from those that are caused
by the $r$-dependence of the polarizability anisotropy. 

Additionally, the interferometric technique may be utilized to probe
local electric fields, such as the radio-frequency fields at different
positions within  a (linear) Paul trap. Obviously, the rotational
Ramsey interferometer will work as
well for other molecular ion species, provided the laser light does
not drive resonant transitions.

\begin{acknowledgements}
We would like to thank Rosario Gonz\'alez-F\'erez 
and Daniel Reich for fruitful discussions. Financial
support by the COST Action MP1001 IOTA, the Danish Ministry of
Higher Education and Science through the Sapere Aude program and 
the European Commission under the Seventh Framework Programme FP7
GA 607491 COMIQ is gratefully acknowledged.  
\end{acknowledgements}


\end{document}